\documentclass[trackchanges,twocolumn]{aastex701}

\hypersetup{citecolor=blue, filecolor=blue, urlcolor=blue}

\usepackage{amsmath}
\usepackage{natbib}
\usepackage{booktabs} 
\usepackage{comment}
\usepackage{adjustbox}
\usepackage{multirow}

\newcommand{\ksf}[1]{\textcolor{teal}{ksf: #1}}
\newcommand{\ace}[1]{\textcolor{red}{ace: #1}}

\begin{document}

\title{Quasar clustering and duty cycle measurements at $0\leq z\leq 4$ with the Gaia-unWISE Catalog 
}

\author[0009-0004-3496-6150]{Mariona Giner Mascarell}
\affiliation{Department of Physics, University of Richmond, Richmond, Virginia, VA 23173, USA}
\affiliation{Department of Physics, Massachusetts Institute of Technology, Cambridge, MA 02139, USA}
\email[show]{mariona.ginermascarell@richmond.edu}  

\author[0000-0003-2895-6218]{Anna-Christina Eilers} 
\affiliation{Department of Physics, Massachusetts Institute of Technology, Cambridge, MA 02139, USA}
\affiliation{MIT Kavli Institute for Astrophysics and Space Research, Massachusetts Institute of Technology, Cambridge, MA 02139, USA}
\email{eilers@mit.edu}

\author[0000-0001-8764-7103]{Kate Storey-Fisher} 
\affiliation{Kavli Institute for Particle Astrophysics and Cosmology, Stanford University, 452 Lomita
Mall, Stanford, CA 94305, USA}
\email{kstoreyf@stanford.edu} 


\begin{abstract}
We measure the two-point correlation function of a uniformly selected, all-sky sample of $\sim$1.3 million quasars with magnitudes $G\leq20.5$ from the \textit{Gaia}–unWISE Quasar Catalog (Quaia) over the redshift range $0 \leq z \leq 4$ to trace the evolution of the quasar clustering strength across cosmic time. We find a steady increase in the correlation length $r_0$ with redshift, i.e. $r_0 = 6.8 \pm 0.2\,h^{-1}\mathrm{Mpc}$ at $0 \leq z < 1$, $r_0=8.0 \pm 0.2\,h^{-1}\mathrm{Mpc}$ at $1 \leq z < 2$, $r_0=10.8 \pm 0.2\,h^{-1}\mathrm{Mpc}$ at $2 \leq z < 3$, and $r_0=13.9 \pm 1.2\,h^{-1}\mathrm{Mpc}$ at $3 \leq z < 4$, and slopes consistent with $\gamma \approx 2$. Our measurements suggest a slightly weaker clustering signal at $z>3$ than previous studies, and thus we find a smooth, monotonic rise in clustering strength. Using a bias-halo mass relation and a step-function for the halo occupation distribution, we infer characteristic minimum halo masses of quasar hosts of $\log_{10}(M_{\mathrm{min}}/M_\odot) \approx 12.8$ across all redshifts. Combining these with the observed quasar number densities yields duty cycles that rise from $f_{\mathrm{duty}} \approx 2\%$ to $\approx 7\%$ with increasing redshift, corresponding to integrated quasar lifetimes of $t_{\rm QSO}\sim10^8$~years. These results suggest that both the characteristic halo mass of active quasars and their typical lifetimes have remained remarkably stable over more than $12 \sim$ Gyr of cosmic time, implying a self-regulated growth process largely independent of epoch.
\end{abstract}

\keywords{}

\section{Introduction} \label{sec:intro}

Nearly all galaxies in the local universe are known to host central supermassive black holes (SMBHs). Well-established empirical correlations between SMBH mass and their host properties (e.g., bulge mass and velocity dispersion) point towards coupled growth between the SMBHs and their hosts \citep{KormendyHo2013, McConnellMa2013}. The classical Soltan argument \citep{Soltan1982} links the time galaxies shine as luminous quasars and during which the bulk of the black hole growth is believed to occur to the local SMBH mass density \citep[see also][]{YuTremaine2002}, suggesting that a substantial fraction of SMBH assembly occurs during radiatively efficient, UV-luminous quasar episodes. 

Quasars are expected to trace the most massive peaks of the cosmic density field, serving as powerful probes between baryonic mass and the underlying dark matter density field \citep[e.g.][]{White2008,Hopkins2006}. The quasars' clustering amplitude and evolution encode how SMBHs populate dark matter halos and how accretion onto the SMBHs is triggered across cosmic time. In this work, we will study the clustering of quasars focusing on the redshift interval $0 \leq z \leq4$, which spans the rise and decline of cosmic star formation and quasar activity \citep{MadauDickinson2014}, providing a key window into both the assembly of massive galaxies and the growth of their central black holes. 

By measuring the clustering strengths of quasars via the auto-correlation function we will determine their correlation length scale $r_0$, and the quasars' minimum host dark matter halo mass $M_{\rm min}$. 
Our cosmological model $\Lambda$CDM provides a framework to link the host halo mass to their number density, and thus a comparison between the number densities of observed quasars to the number densities of their host dark matter halos inferred from clustering studies will also allow us to constrain the quasars' duty cycle, which is the fraction of cosmic time that SMBHs grow as luminous quasars 
\citep{MartiniWeinberg2001,ConroyWhite2013}. In particular, we can estimate the duty cycle as
\begin{equation}
  f_{\mathrm{duty}}\simeq\frac{t_{\rm QSO}}{t_{\rm H}(z)}
  \;\simeq\;
  \frac{n_{\mathrm{QSO}}}
       {n_{\mathrm{DM}}(>M_{\min})}
  \label{eq:fduty}
\end{equation}
where $t_{\rm QSO}$ denotes the typical quasar lifetime, $t_{\rm H}(z)$ is the Hubble time at redshift $z$, $n_{\mathrm{QSO}}$ is the observed comoving number density of quasars, and $n_{\mathrm{DM}}(>M_{\min})$ is the comoving number density of dark matter haloes above the characteristic minimum host mass inferred from the clustering analysis.

Previous clustering analyses have established this framework for quasars across a wide range of redshifts and luminosities. At moderate redshift ($0.8 \leq z \leq 2.1$), the Sloan Digital Sky Survey (SDSS) Data Release~5 (DR5;~\citealt{Shen2007}) provided one of the first large, uniform quasar clustering measurements, based on a sample of $\sim$30{,}000 quasars. Later SDSS releases extended these analyses with larger samples and improved redshift coverage (e.g.,~\citealt{White2012,Laurent2017}), but the derived clustering amplitudes remain broadly consistent with DR5 results. These studies find correlation lengths of $r_0 = 5.45^{+0.35}_{-0.45}\,h^{-1}\mathrm{Mpc}$, corresponding to characteristic halo masses of order $10^{12}\,h^{-1}M_\odot$ and duty cycles of a few percent. At higher redshifts ($2.2 \leq z \leq 2.8$), \citet{Eftekharzadeh2015} measured $r_0 = (8.12 \pm 0.22)\,h^{-1}\mathrm{Mpc}$, consistent with halo masses of $\sim2\times10^{12}\,h^{-1}M_\odot$ and similarly low duty cycles, with little dependence on luminosity.
However, at higher redshifts of $3\lesssim z\lesssim4$, clustering analyses of SDSS quasars have revealed a pronounced increase in bias with redshift \citep{Shen2007, White2012, Eftekharzadeh2015, Arita2023}. For instance, \citet{Shen2007} found $r_0 = 16.9 \pm 1.7\,h^{-1}\mathrm{Mpc}$ at $2.9 \le z \le 3.5$ and $r_0 = 24.3 \pm 2.4\,h^{-1}\mathrm{Mpc}$ at $z \ge 3.5$, corresponding to halo masses of $\sim\!2$--$6\times10^{12}\,h^{-1}M_\odot$ and quasar lifetimes of several hundred Myr, implying duty cycles approaching unity.

Beyond $z\gtrsim6$, quasar auto-correlation measurements become increasingly challenging due to the low number density of luminous quasars. Using a faint ($M_{1450}\gtrsim-25$), but more numerous population of quasars a first attempt of the auto-correlation measurement at $z\sim 6$ has recently been reported by the Subaru High-z Exploration of Low-luminosity Quasars (SHELLQs) Collaboration \citep{Arita2023}, indicating host dark matter halo masses similar or slightly higher than those found at lower redshifts, but uncertainties are large. However, while auto-correlation measurements of quasars at these high redshifts remain challenging, complementary quasar-galaxy cross-correlation measurement is still viable. Recent measurements of quasar–galaxy cross-correlations leveraging data from the James Webb Space Telescope (JWST) reveal similar minimum host masses of $\log_{10}(M_{\rm halo,min}/M_\odot)=12.43^{+0.13}_{-0.15}$ to those in the lower redshift universe, and extremely short UV-bright duty cycles $f_{\rm duty}\lesssim 1\%$, suggesting only brief luminous black hole growth phases \citep{Eilers2024, Pizzati2024}. At even earlier cosmic times, the picture becomes less clear; for example, a recent study of two quasar fields at $z\approx7.3$ suggests a decrease in host halo mass down to $\log_{10}(M_{\rm halo,min}/M_\odot)\sim 11.6\pm0.6$ and duty cycles of $f_{\rm duty}\sim 0.1\%$ \citep{Schindler2025}, potentially indicating a non-monotonic redshift evolution. 
Despite these advances, current measurements are highly heterogeneous in their sample selection, making direct comparisons across redshift challenging. Thus, a homogeneous clustering  analysis is essential to understand how quasar activity and host halo properties evolve across cosmic history to the present day. 

In this work, we leverage the \textit{Gaia}-unWISE Quasar Catalog (``Quaia’’; \citealp{StoreyFisher2024}), an all-sky, homogeneously observed sample of $\sim$1.3 million quasars. 
Its combination of full-sky coverage, relatively uniform depth, and a tractable selection function provide an unprecedented data set for measuring the quasar clustering signal. 
We measure the projected two-point correlation function $w_p(r_p)$ in four redshift bins between $0\leq z\leq4$, and derive the real-space correlation length $r_0$ and slope $\gamma$. We then use a halo-bias framework \citep{Tinker2008,Tinker2010} to infer the corresponding minimum host dark matter halo mass $M_{\rm min}$ and derive the quasars' duty cycle $f_{\rm duty}$ as a function of redshift. 

This paper is structured as follows. In Section \ref{sec:data}, we describe the Quaia catalog and the construction of our quasar sample, along with the selection functions and catalogs of randomly distributed quasars used to correct for observational biases. Section \ref{sec:corrfunc} presents our methodology for computing the projected two-point correlation function, and measuring the correlation lengthscales. We also contrast our clustering results with previous studies in the literature. In Section \ref{sec:dmduty}, we infer the characteristic host dark-matter halo masses of Quaia quasars and estimate their duty cycles by combining our clustering measurements with quasar number densities. Finally, Section \ref{sec:discussion} summarizes our main findings and discusses their implications for quasar evolution and large-scale structure at high redshift. 
Throughout, we adopt a \citet{Planck2018} cosmology:
$
\Omega_m = 0.315,\quad
\Omega_\Lambda = 0.685,\quad
\Omega_b = 0.049,\quad
\Omega_c = 0.265,\quad
\Omega_r \simeq 9\times10^{-5},\quad
\Omega_k \simeq 0.
$

\section{Data}
\label{sec:data}

\subsection{Quaia: Gaia's quasar sample}

We use the all-sky quasar catalog Quaia \citep{StoreyFisher2024}, constructed from a parent sample of \textit{Gaia}~DR3 quasar candidates \citep{GaiaDR3QSOC} cross-matched with \textit{unWISE} infrared photometry \citep{Lang2016,Meisner2019}. The Quaia quasar sample is described in detail in \citet{StoreyFisher2024}, but we will summarize the properties here briefly. 

\begin{figure}
  \centering
  \includegraphics[width=\columnwidth]{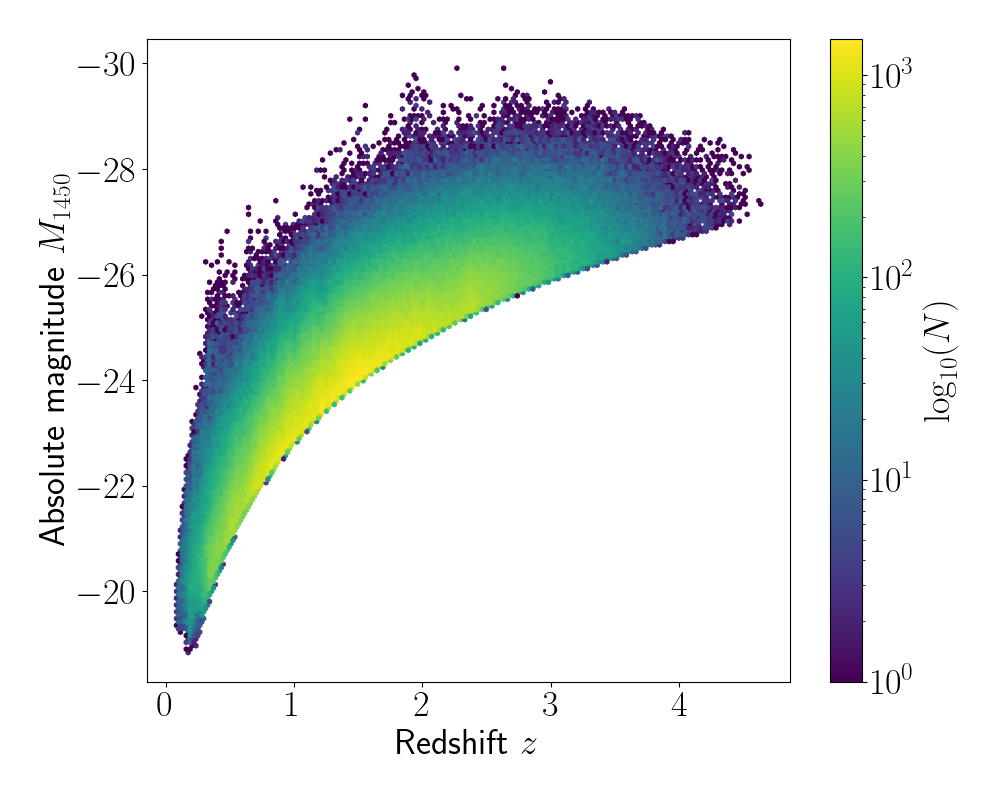}
  \caption{
Two-dimensional histogram of the Quaia quasar sample as a function of redshift and absolute magnitude $M_{1450}$. 
}

  \label{fig:sample}
\end{figure}

The parent \textit{Gaia} DR3 quasar candidate sample contains $6{,}649{,}162$ sources with low-resolution BP/RP spectra (330–1050\,nm; $R\!\sim\!30$–100) and pipeline redshift estimates (“QSOC”), but was designed for completeness over purity (estimated purity $\sim\!52\%$; \citealt{GaiaDR3QSOC}). \citet{StoreyFisher2024} decontaminated this Gaia parent set by applying a proper-motion cut and color cuts in \textit{Gaia} matched with unWISE colors. This procedure reduced the known contaminants by $\sim\!4\times$ while removing only $\sim\!1.2\%$ of SDSS-confirmed quasars \citep{Ahumada2020}. The final Quaia sample used in this analysis is limited to magnitudes $G<20.5$ and contains $1{,}295{,}502$ quasars. We convert Gaia's $G$-band magnitude to an absolute magnitude at rest-frame $1450$~{\AA}, $M_{1450}$, by simply applying a K-correction assuming a power-law continuum with spectral index $\alpha_\nu=-0.61$ \citep{Lusso2015}, and show the sample in Fig.~\ref{fig:sample}.  

Additionally, the redshift estimates are improved in Quaia by training a $k$-nearest-neighbors model on SDSS spectroscopic labels \citep{Ahumada2020} using \textit{Gaia}+unWISE colors and the “QSOC” redshift estimates from Gaia DR3 as features. For the $G<20.5$ sample the catastrophic outlier fraction is reduced to $9\%$ ($16\%$) at $|\Delta z/(1+z)|>0.2$ ($0.1$), which is a $\sim\!3\times$ ($\sim\!2\times$) improvement over the raw “QSOC” values. The typical estimated redshift precision for the quasar sample with $G<20.5$ is $\Delta z/(1+z)\!=\!0.008$ \citep{StoreyFisher2024}.

\textbf{Purity/completeness.} Quaia is constructed to maximize homogeneity and controllable selection rather than to optimize a single global “purity” number. Empirically, the decontamination step reduces known contaminants by $\sim\!4\times$ while retaining $>\!98\%$ of SDSS quasars; the residual spatial completeness is modeled with an all-sky selection-function map built via Gaussian-process regression on dust, source-density, and survey-scan templates, with residuals consistent with homogeneous noise away from the Galactic plane \citep{StoreyFisher2024}.

\textbf{Volume and comparison to other surveys.} Despite a lower number density than other optical+IR surveys, Quaia spans a larger effective comoving volume than SDSS DR16Q and the eBOSS quasar sample when matched in magnitude: for $G<20.5$, $V_{\rm eff}\!\approx\!6.5\,h^{-3}{\rm\,Gpc}^3$, nearly $2\times$ DR16Q and $>\!2\times$ eBOSS \citep{Lyke2020,Myers2015}; over the common span $0.8 \leq z \leq 2.2$, the “spanning volume” reaches $\sim\!1.44\times10^2\,h^{-3}{\rm\,Gpc}^3$ \citep{StoreyFisher2024}.

For our main analysis, we split the $G<20.5$ quasars into four redshift bins: 

\[
\begin{aligned}
0.0 \leq z < 1.0 &:\; N = 327{,}195, \\
1.0 \leq z < 2.0 &:\; N = 648{,}095, \\
2.0 \leq z < 3.0 &:\; N = 283{,}958, \\
3.0 \leq z < 4.0 &:\; N = 35{,}466. 
\end{aligned}
\]

\subsection{Selection functions and random catalogs}
\label{subsec:selfunc_randoms}

To estimate the clustering of quasars, we require a catalog of randomly distributed quasars that accurately reproduces both the angular and redshift selection functions of our data set. Measuring the quasars’ auto-correlation function therefore depends on a well-modeled selection function, which is used to create the catalogs of randomly distributed quasars assuming no clustering signal is present. Thus, these random catalogs provide the baseline against which we measure the excess clustering of the real quasar sample. To construct the selection functions for our quasar sample, we adopt the data-driven framework developed in \citet{StoreyFisher2024}, which learns how survey systematics modulate the observed quasar number density on the sky. The regression is performed on \textsc{HEALPix} pixels \citep{Gorski2005} (NSIDE=64) using a compact set of physically motivated templates. These include the Corrected Schlegel, Finkbeiner, and Davis (CSFD) dust reddening map \citep{Chiang2023}, which removes extragalactic cosmic infrared background (CIB) contamination from the classic SFD map \citep{Schlegel1998}; stellar density maps constructed from \textit{Gaia} source counts with $18.5<G<20$; unWISE source-density maps \citep{Lang2016}; the \textit{Gaia} scanning law and crowding proxy $M_{10}$, defined as the median $G$ magnitude of sources with $\leq10$ transits from \texttt{GaiaUnlimited} \citep{GaiaUnlimited}, which encodes the scanning law and crowding effects; the unWISE scanning law, given by the mean number of single-exposure W1 images contributing to each coadded image \citep{Meisner2019}; and localized versions of the stellar and unWISE density maps restricted to wide regions around the Magellanic Clouds to capture their distinct contamination patterns.

Following \citet{StoreyFisher2024}, we regress the (log) quasar counts per pixel on the mean-subtracted templates using a Gaussian-process model with a squared-exponential kernel, which captures the nonlinear response of the number density to the systematics. The prediction is then converted into a relative completeness by normalizing to the mean predicted density in a set of ``clean'' pixels, which are defined as those with low extinction, low stellar and unWISE counts, high $M_{10}$, and high unWISE coadds. This normalization ensures values $\leq1$ and avoids sensitivity to a single extremal pixel.

We use selection-function maps and random catalogs at NSIDE=64 for the $G<20.5$ sample. The random catalogs are generated by Poisson sampling a uniform sky and down-weighting by the selection function, and are provided at approximately $100\times$ the density of the data. 

For redshift-split analyses, we adopt the corresponding selection functions and assign redshifts to objects in the random catalog by sampling from the empirical $p(z)$ of each bin. This procedure overall captures the Galactic extinction for quasars close to the plane very well. However, we note that for our analysis in the highest redshift bin, i.e.\ $3\leq z\leq4$, we apply a further cut based on Galactic latitude, excluding quasars close to the Galactic plane at $|b|<30^\circ$, where modeling the dust distribution and its associated extinction in the optical bands remains uncertain, in order to mitigate contamination as we otherwise observe nonphysical clustering effects on large scales. 

\begin{figure*}
  \centering
  \includegraphics[width=\textwidth]{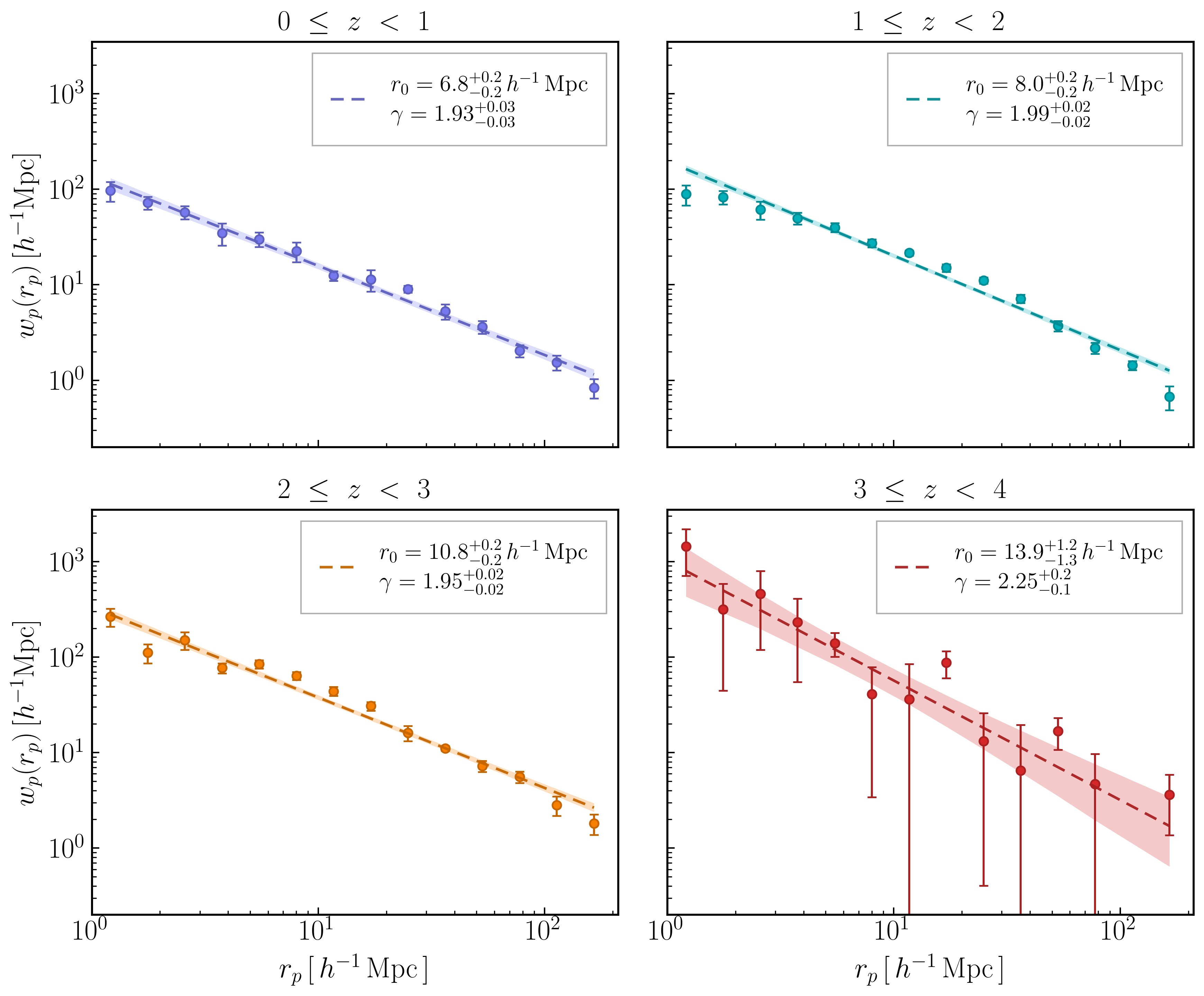}
  \caption{The projected auto-correlation function for the four redshift bins is shown with jackknifing errors for the four redshift bins, with errors estimated using jackknife resampling. The dashed lines and shaded areas show the best power-law fits with $2 \sigma$ uncertainties. 
  The legend shows the best-fit parameter values of the power-law model fit to the data. 
    }
  \label{fig:wp_all}
\end{figure*}

\section{Quasar correlation function} \label{sec:corrfunc}

\subsection{Auto-correlation function and scale length measurements}
\label{sec:methods:correlation}

In this section, we present measurements of the projected two-point correlation function of quasars in distinct redshift intervals between $0\leq z < 4$. We apply the Landy–Szalay estimator \citep{LandySzalay} as shown in Eqn.~\ref{eq:ls}, which is subsequently integrated along the line of sight to yield the projected correlation function. We then fit these projected correlation functions with a power-law, to study the clustering properties of the quasar population.

For each redshift bin we generate a corresponding selection function to construct a catalog of quasars randomly distributed on the sky. The Landy–Szalay estimator is defined as
\begin{equation}
\xi(r) = \frac{\langle\rm DD \rangle - 2\langle DR \rangle + \langle RR \rangle}{\langle \rm RR \rangle},
\label{eq:ls}
\end{equation}
where $\langle \rm DD \rangle$, $\langle\rm DR \rangle$, and $\langle \rm RR \rangle$ denote the normalized counts of data–data, data–random, and random–random pairs, respectively. These pairs are counted as a function of distance, $r = \sqrt{r_p^2 + \pi^2}$, where $r_p$ and $\pi$ denote the separations perpendicular and parallel to the line of sight, respectively.

We compute the projected two-point correlation function, $w_p(r_p)$, by integrating the redshift-space correlation function, $\xi(r_p,\pi)$, along the line of sight:  

\begin{equation}
    w_p(r_p) = 2 \int_{0}^{\pi_{\max}} \xi(r_p,\pi)\, d\pi. 
\end{equation}
This projection suppresses the impact of redshift-space distortions, yielding an estimator that more directly reflects the underlying real-space clustering of quasars. In our analysis, we adopt $\pi_{\max} = 100 \, h^{-1}\,\mathrm{Mpc}$ for the line-of-sight integration, a choice commonly used in quasar clustering studies \citep[e.g.,][]{Shen2007,White2012,Eftekharzadeh2015,Arita2023}. We compute $w_p(r_p)$ in 14 logarithmically spaced transverse-separation bins spanning $r_p = 1$--$200 \, h^{-1}\,\mathrm{Mpc}$ (corresponding to a bin width of $\Delta \log r_p \simeq 0.16$).

The statistical uncertainties on $w_p(r_p)$ are estimated using delete-one jackknife resampling within each redshift bin. We divide both the quasars and the random catalog into ten equal-width right ascension (RA) stripes covering the sky region of the bin. The projected correlation function is then recomputed ten times, each time omitting one RA stripe. Let $w_p^{(k)}(r_p)$ denote the measurement obtained when stripe $k$ is excluded, and let $\bar{w}_p(r_p)$ be the mean over all jackknife realizations. The jackknife covariance matrix is computed using

\begin{align}
\label{covariance}
C_{ij} = \frac{N-1}{N}\sum_{k=1}^{N}
&\left[w_p^{(k)}(r_{p,i}) - \bar{w}_p(r_{p,i}) \right] \nonumber \\ 
&\left[w_p^{(k)}(r_{p,j}) - \bar{w}_p(r_{p,j}) \right], 
\end{align}
where $N=10$ is the number of jackknife regions. For visualization and for the tabulated uncertainties, we use only the diagonal elements of this matrix to define the $1\sigma$ error bars,

\[
\sigma_{w_p}(r_{p,i}) = \sqrt{C_{ii}}.
\]
Our measurements of $w_p(r_p)$, along with the corresponding transverse separations $r_p$ and jackknife uncertainties $\sigma_{w_p}$, are shown in Fig.~\ref{fig:wp_all} and listed in Table~\ref{tab:wp_allbins} for each redshift bin.

To interpret the measurements of $w_p(r_p)$, we model the two-point correlation function as a power law of the form  

\begin{equation}
    \xi(r) = \left( \frac{r}{r_0} \right)^{-\gamma},
\end{equation}
where $r_0$ is the correlation length and $\gamma$ is the slope. This yields the following expression for the projected correlation function, which can be derived by integrating along the line-of-sight:

\begin{equation}
    w_p(r_p) = r_p \left( \frac{r_0}{r_p} \right)^{\gamma} 
    \frac{\Gamma\!\left(\tfrac{1}{2}\right)\Gamma\!\left(\tfrac{\gamma - 1}{2}\right)}{\Gamma\!\left(\tfrac{\gamma}{2}\right)} ,
\end{equation}
where $\Gamma$ denotes the Gamma function \citep{Shen2007, DavisPeebles1983}. 

\begin{figure*}
  \centering
  \includegraphics[width=\textwidth]{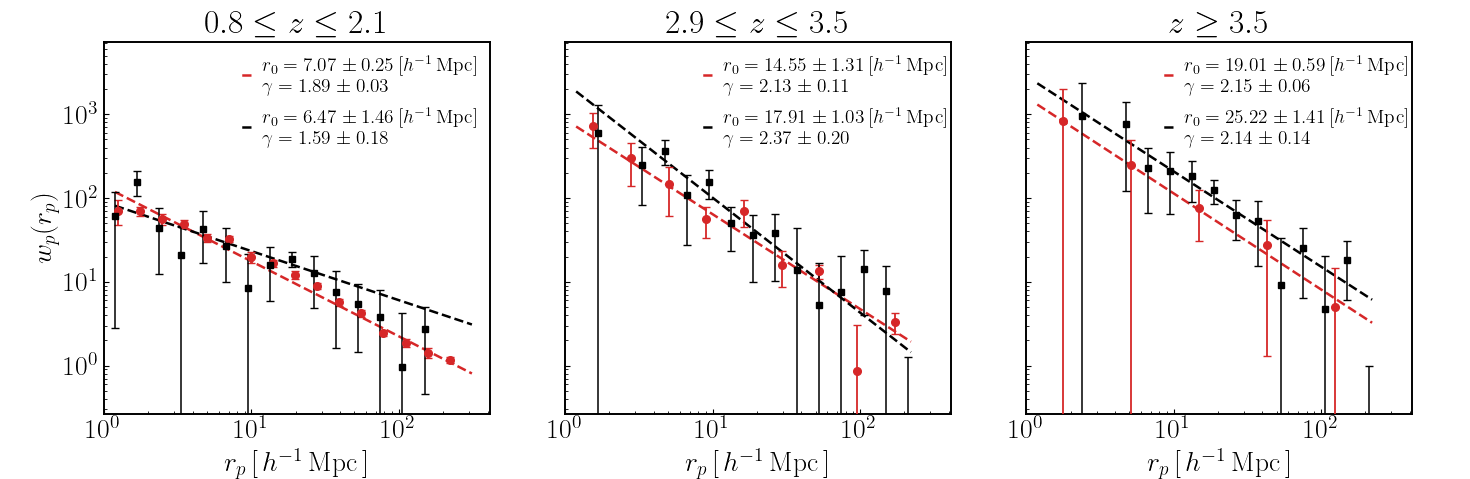}
  \caption{Comparison of the projected auto-correlation function of Quaia sources with SDSS sources analyzed by \citet{Shen2007} for $0.8 \leq z \leq 2.1$, $2.9 \leq z \leq 3.5$, and $ z \geq 3.5$ (their \textit{good} fields). Black measurements indicate results by \citet{Shen2007}, while red data points show the results from this study.}
  \label{fig:shen}
\end{figure*}

We fit this power-law model to the measured $w_p(r_p)$ in each redshift bin to infer the scale length $r_0$ and slope $\gamma$. Parameter estimation uses the ensemble Markov Chain Monte Carlo sampler \texttt{emcee} \citep{emcee} to explore the posterior $p(r_0,\gamma\,|\,\mathbf{w}_p)$. We adopt a Gaussian likelihood with the measured $w_p(r_p)$ values as the data vector and a jackknife-estimated, diagonal covariance (i.e., we use only the jackknife variances $\sigma_{w_p}^2$ for each $r_p$ bin). We impose uniform priors matching of $r_0 \in [1.0,\,50.0]$ and $\gamma \in [1.1,\,3.0]$. We report the posterior medians as fiducial estimates and the 16th--84th percentiles as the corresponding $1\sigma$ credible intervals. The resulting values of $r_0$ and $\gamma$ are reported in Table~\ref{tab:fits}, and the best-fit models are shown in Figure~\ref{fig:wp_all}.

\begin{table*}[htbp]
\setlength{\tabcolsep}{15pt}
\renewcommand{\arraystretch}{1.2}
\centering
\caption{Projected correlation function $w_p(r_p)$ across redshift bins.}
\label{tab:wp_allbins}
\begin{adjustbox}{max width=\textwidth}
\begin{tabular}{ccccc}
\toprule
\multirow{2}{*}{$r_p \,[h^{-1}\mathrm{Mpc}]$} 
& \multicolumn{4}{c}{$w_p(r_p) \pm \sigma_{w_p} [h^{-1}\mathrm{Mpc}]$} \\
\cmidrule(lr){2-5}
& $0 \leq z<1$ & $1 \leq z<2$ & $2 \leq z<3$ & $3 \leq z<4$ \\
\midrule
1.47 & $79.1 \pm 12.8$ & $79.2 \pm 12.2$ & $178 \pm 13.5$ & $677 \pm 517$ \\
3.16 & $37.2 \pm 4.39$ & $52.8 \pm 4.75$ & $97.5 \pm 18.3$ & $326 \pm 171$ \\
6.81 & $22.4 \pm 2.95$ & $31.9 \pm 3.07$ & $67.0 \pm 4.81$ & $45.3 \pm 24.3$ \\
14.7 & $10.7 \pm 2.24$ & $16.3 \pm 1.46$ & $29.7 \pm 2.28$ & $72.7 \pm 23.1$ \\
31.6 & $5.43 \pm 0.76$ & $7.07 \pm 0.75$ & $11.2 \pm 0.76$ & $10.3 \pm 10.0$ \\
68.1 & $1.78 \pm 0.32$ & $2.18 \pm 0.22$ & $4.73 \pm 0.54$ & $4.83 \pm 3.55$ \\
147  & $0.747 \pm 0.168$ & $0.544 \pm 0.155$ & $1.38 \pm 0.36$ & $3.98 \pm 0.85$ \\
316  & $0.313 \pm 0.086$ & $0.156 \pm 0.074$ & $0.285 \pm 0.240$ & $2.49 \pm 1.07$ \\
681  & $0.133 \pm 0.026$ & $0.121 \pm 0.030$ & $0.228 \pm 0.119$ & $1.05 \pm 0.61$ \\
\bottomrule
\label{maintable}
\end{tabular}
\end{adjustbox}
\end{table*}

\begin{table}[ht]
    \caption{Best-fit parameters $r_0$ and $\gamma$ in four redshift bins.}
    \label{tab:fits}
    \setlength{\tabcolsep}{8pt}
    \renewcommand{\arraystretch}{1.2}
    \centering
    \begin{tabular}{ccc}
        \hline
        Redshift range & $\gamma$ & $r_{0}\,[h^{-1}\,\mathrm{Mpc}]$ \\
        \hline
        $0.0 \leq z < 1.0$ & $1.93 \pm 0.03$ & $6.8 \pm 0.2$ \\ 
        $1.0 \leq z < 2.0$ & $1.99 \pm 0.02$ & $8.0 \pm 0.2$ \\
        $2.0 \leq z < 3.0$ & $1.95 \pm 0.02$ & $10.8 \pm 0.2$ \\
        $3.0 \leq z < 4.0$ & $2.25 \pm 0.02$ & $13.9 \pm 1.2$ \\
        \hline
    \end{tabular}
\end{table}

\subsection{Comparison to previous work}
\label{compareshen}

We compare our measurements to the quasar sample of comparable brightness analyzed by \citet{Shen2007}, who studied the large-scale clustering of $\sim45{,}000$ optically selected SDSS DR5 quasars spanning $0.8 \le z \le 4.0$ and $M_i < -22$ (K-corrected to $z=2$) over $\sim4{,}000~\mathrm{deg}^2$ of sky. Their sample, drawn from the SDSS color-based quasar selection algorithm, probes bright unobscured quasars with redshifts and luminosities approximately matching those of our Gaia–unWISE sample. Thus, both datasets trace a similar population of luminous, optically selected quasars over comparable redshift ranges.


To ensure a consistent comparison, we adopt the same redshift intervals used in the previous study, i.e. $0.8 \le z < 2.1$, $2.9 \le z < 3.5$, and $z \ge 3.5$, for which we find $N=814,858$, $N=41,745$, and $N=6,783$ quasars, respectively. Note that we do apply the same cut in Galactic latitude as before, i.e. $|b|>30^\circ$, for the latter two redshift bins. For this part of the analysis, we recompute the \textit{Quaia} selection functions and random catalogs within these bins, ensuring that the projected correlation functions $w_p(r_p)$ are measured under with an identical redshift selection. This allows a direct comparison of the clustering amplitude and slope between \textit{Quaia} and the SDSS spectroscopic results of \citet{Shen2007}. Figure~\ref{fig:shen} shows the projected correlation functions reported by \citet{Shen2007} in black (for their \textit{good} fields, which is a subset of \textit{all} quasar fields analyzed in their work that satisfy a set of quality criteria on the photometry, see their \S~$2.2$ for details) alongside our measurements shown in red. 
Overall, we find excellent agreement between the two analyses at lower redshift. For the $0.8 \leq z < 2.1$ bin, \citet{Shen2007} reports $r_0=6.47\pm1.55\,h^{-1}\,\mathrm{Mpc}$ and $\gamma=1.58\pm0.20$, which are slightly lower than our $r_0=7.07\pm 0.25 \,h^{-1}\,\mathrm{Mpc}$ and $\gamma=1.89\pm 0.03$. However, at $2.9 \leq z \leq 3.5$, \citet{Shen2007} finds $r_0=17.91\pm1.51 \,h^{-1}\,\mathrm{Mpc}$ and $\gamma=2.37\pm0.29$ while we measure slightly lower values of $r_0=13.05\pm 1.9\,h^{-1}\,\mathrm{Mpc}$ and $\gamma=2.06\pm0.17$. At even higher redshift, at $z \geq 3.5$, this discrepancy increases to $r_0=25.22\pm 2.5\,h^{-1}\,\mathrm{Mpc}$ and $\gamma=2.14\pm0.24$ reported from the SDSS quasars, while our analysis yields $r_0=19.01\pm 0.6 \,h^{-1}\,\mathrm{Mpc}$ and $\gamma=2.15\pm0.06$.
Note, that our results are consistent with measurements by \citet{Shen2007} when \textit{all} fields are taken into account (instead of only their \textit{good} fields) and when they fix the slope to $\gamma=2.0$, in which case they obtain $r_0=14.79\pm2.12\,h^{-1}\,\mathrm{Mpc}$ for $2.9\leq z\leq 3.5$ and $r_0=20.68\pm2.52 \,h^{-1}\,\mathrm{Mpc}$ at $z\geq 3.5$.

\section{Host Dark Matter Halo Masses and Quasar Duty Cycle}
\label{sec:dmduty}

\subsection{Estimating the minimum host dark matter halo mass}
\label{sec:methods:ndm}

In the framework of hierarchical structure formation, quasars are assumed to reside in dark matter haloes, and their clustering properties therefore encode information about the typical halo masses that host them. Within the $\Lambda$CDM cosmological model, the strength of clustering on large scales (quantified by the correlation length $r_{0}$) reflects the typical separation scale over which quasars remain correlated. In other words, $r_{0}$ marks the distance at which the real-space two-point correlation function $\xi(r)$ falls to unity, meaning that a pair of quasars separated by $r_0$ is twice as likely to be found together as if they were randomly distributed. A larger $r_{0}$ therefore corresponds to stronger clustering and indicates that quasars inhabit rarer, more strongly biased environments. More strongly clustered quasars correspond to rarer, more massive haloes, while weaker clustering reflects residence in lower-mass haloes. Concretely, the observed $r_{0}$ can be mapped onto a minimum host-halo mass $M_{\min}$ by requiring consistency between the quasar number density and the cumulative abundance of dark matter haloes, $n_{\mathrm{DM}}(>M_{\min})$.

We perform this conversion using the publicly available \textsc{halomod} \texttt{TracerHaloModel} framework \citep[e.g.][]{Hearin2017}, which self-consistently incorporates the halo mass function and halo bias relation to connect the measured clustering amplitude with a characteristic halo mass. In our configuration, \textsc{halomod} adopts the halo mass function of \citet{Tinker2008} together with the large-scale halo bias model of \citet{Tinker2010}, which enter through the two-halo term used to predict the large-scale real-space correlation function. In practice, we compute the model correlation function $\xi(r\!\mid\!M_{\min})$ for a grid of $M_{\min}$ values, extract the corresponding correlation length $r_0(M_{\min})$ from the condition $\xi(r_0)=1$, and invert this mapping to obtain the value of $M_{\min}$ that reproduces the observed $r_0$ in each redshift bin. Tracer occupation is set to unity above a sharp mass threshold (a step-function, or “constant-HOD”, model), leaving the minimum halo mass $M_{\min}$ as the single free parameter that links halo mass to the observed quasar clustering.

For each effective redshift 
$z_{\mathrm{eff}} = \{0.5,\,1.5,\,2.5,\,3.5\}$, 
we tabulate the function $r_{0}(M_{\min})$ on a grid of minimum halo masses spanning 
$9 \le \log_{10}(M_{\min}/h^{-1}M_\odot) \le 15$. 
For every trial $M_{\min}$ we (i) populate haloes above that threshold, (ii) compute the real-space halo auto-correlation function $\xi_{\mathrm h}(r)$ over 
$0.1$--$200\,h^{-1}\,\mathrm{Mpc}$, and (iii) record the scale where $\xi_{\mathrm h}=1$. 
The resulting monotonic $r_{0}(M_{\min})$ relation is then inverted and evaluated at the observed median $r_0$ (as well as its 16th and 84th percentiles), yielding the corresponding $M_{\min}$ values with asymmetric $1\sigma$ uncertainties. 

\begin{figure}
  \centering
  \includegraphics[width=\columnwidth]{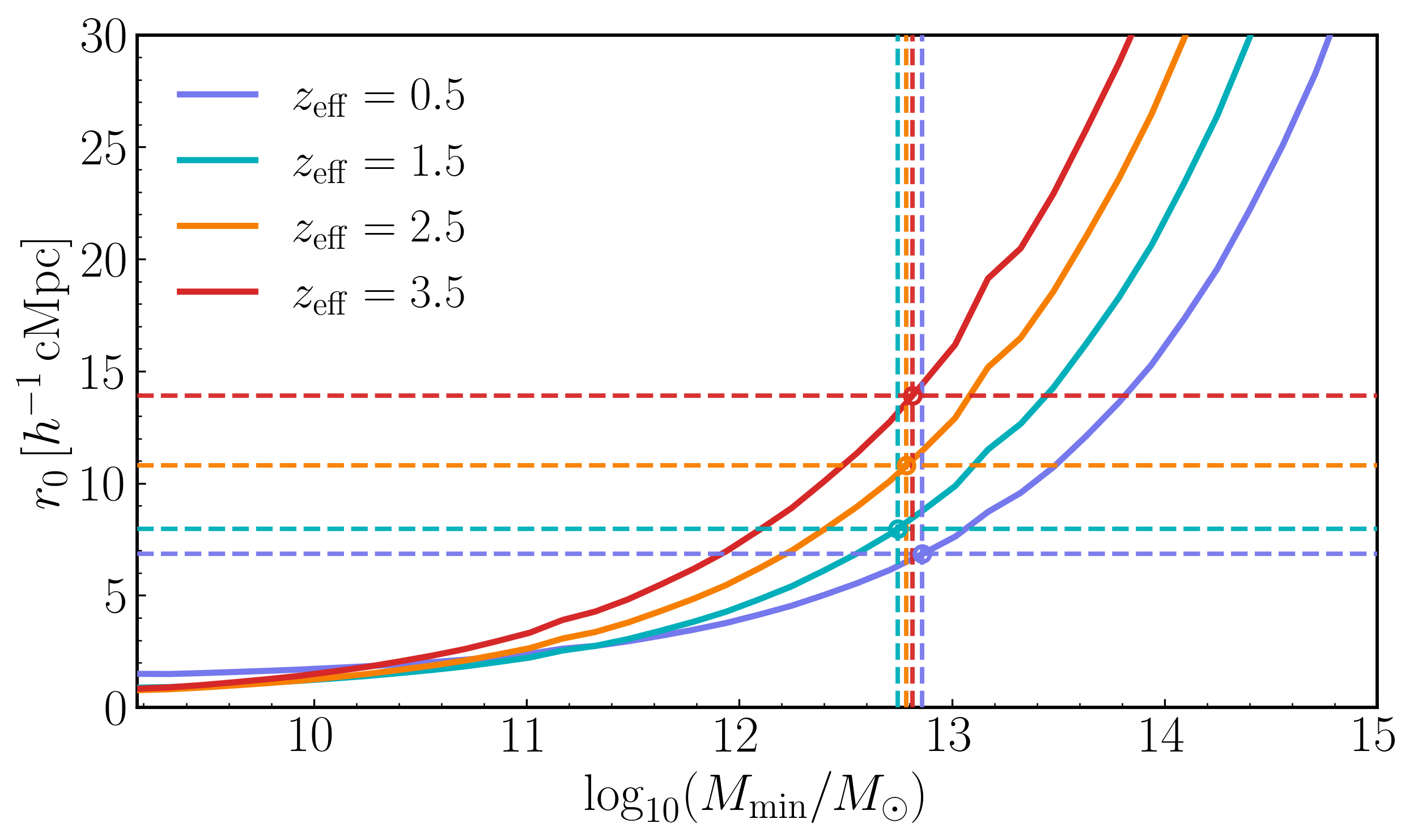}
  \caption{Minimum dark matter halo mass vs $r_0$ comparison. Horizontal dashed lines represent measured values through auto-correlation fits; vertical lines are the predicted mass for each measurement. }
  \label{fig:dm}
\end{figure}

The results are shown in Figure~\ref{fig:dm},  illustrating this mapping between the measured clustering amplitude and the inferred minimum host-halo mass. Although the observed correlation lengths $r_0$ vary substantially with redshift, the corresponding $M_{\min}$ values lie within a relatively narrow range of a few $\times10^{12}\,h^{-1}M_\odot$. This visual trend reflects the redshift evolution of halo bias: at higher redshift, halos of a given mass are rarer and therefore more strongly clustered. As a result, a smaller $r_0$ at lower redshift can correspond to roughly the same halo mass as a larger $r_0$ at high-$z$. The near convergence of the curves highlights that luminous quasars occupy halos of roughly constant characteristic mass across cosmic time.

The cumulative number density of halos more massive than the
threshold is
\[
  n_{\mathrm{DM}}(>M_{\min}) =
  \int_{M_{\min}}^{\infty}
  \frac{{\rm d}n}{{\rm d}M}\,{\rm d}M ,
\]
where ${\rm d}n/{\rm d}M$ is the mass function \citep{Tinker2010}
evaluated at $z_{\mathrm{eff}}$.
Integrating at the lower, median, and upper $M_{\min}$ values
propagates the clustering uncertainty into asymmetric errors on
$n_{\mathrm{DM}}$.

\subsection{Quasar Number Density and Duty Cycle}
\label{sec:methods:duty}

As quasars temporarily and randomly subsample their host dark matter haloes, we can estimate their duty cycle as established in Eqn.~\ref{eq:fduty}. The first expression emphasizes the interpretation of $f_{\mathrm{duty}}$ as the fraction of a Hubble time during which a halo is observed in a luminous quasar phase, while the second expression provides a practical estimator based on the fraction of halos more massive than $M_{\min}$ that are currently hosting a luminous quasar. 

We estimate the quasars' number density $n_{\rm QSO}$ as follows. The \textsc{HEALPix} $N_{\mathrm{side}}=64$ map 
encodes the relative completeness of every sky pixel.
If $c_p$ and $\Omega_p$ denote the weight and solid angle in steradians of pixel $p$, respectively, the survey’s effective sky fraction is

\begin{equation}
  f_{\mathrm{sky,eff}}
  \;=\;
  \frac{1}{4\pi}\,
  \sum_{p} c_p\,\Omega_p. 
  \label{eq:fsky}
\end{equation}

For each redshift slice, the comoving shell volume $V_{\mathrm c}$ that would be
enclosed on the full sky is computed via
$V_{\mathrm{shell}} =
 V_{\mathrm c}(z_{\max}) - V_{\mathrm c}(z_{\min})$.
Multiplication by the factor in Eqn.~\eqref{eq:fsky}
yields the survey’s effective volume, i.e.\
\begin{equation}
  V_{\mathrm{eff}}
  \;=\;
  f_{\mathrm{sky,eff}}\,V_{\mathrm{shell}}. 
  \label{eq:veff}
\end{equation}
Then, given the number of quasars in the given redshift shell as $N_{\mathrm Q}$, the mean quasar space density follows as
\begin{equation}
  n_{\mathrm{QSO}}
  \;=\;
  \frac{N_{\mathrm Q}}{V_{\mathrm{eff}}} ~. 
  \label{eq:nqso}
\end{equation}

\begin{table}[t]
\caption{
Minimum host-halo mass $M_{\min}$ and quasar duty cycle $f_{\rm duty}$ in each redshift bin.
}
\centering
\setlength{\tabcolsep}{8pt}
\renewcommand{\arraystretch}{1.2}
\begin{tabular}{c c c}
\hline
Redshift range & $\log_{10}(M_{\min}/M_\odot)$ & $f_{\rm duty}$ \\
\hline
$0.0 \le z < 1.0$ & $12.86^{+0.04}_{-0.05}$ & $0.018^{+0.002}_{-0.002}$ \\
$1.0 \le z < 2.0$ & $12.74^{+0.02}_{-0.02}$ & $0.022^{+0.001}_{-0.001}$ \\
$2.0 \le z < 3.0$ & $12.79^{+0.03}_{-0.03}$ & $0.056^{+0.008}_{-0.007}$ \\
$3.0 \le z < 4.0$ & $12.82^{+0.10}_{-0.12}$ & $0.071^{+0.077}_{-0.034}$ \\
\hline
\end{tabular}

\label{tab:Mmin_fduty}
\end{table}

\begin{figure*}
  \centering
  \includegraphics[width=\textwidth]{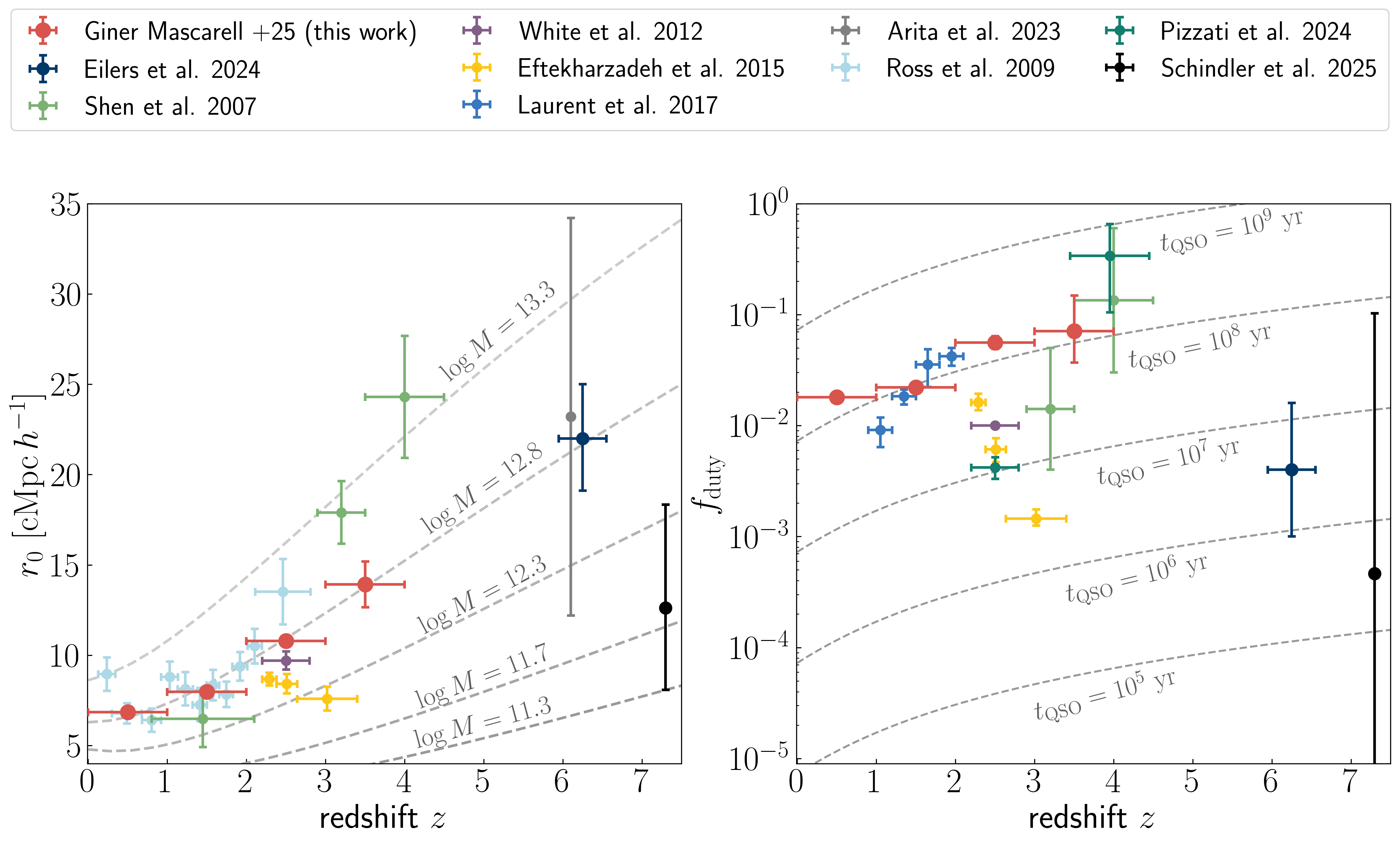}
  \caption{Redshift evolution of the quasars’ auto-correlation scale length (left) and the quasars’ duty cycle (right). Our measurements are shown in red, while results from previous studies are shown in different colors \citep{Shen2007, Ross2009, White2012, Eftekharzadeh2015, Laurent2017, Arita2023, Eilers2024, Pizzati2024, Schindler2025}. Dashed lines in the left panel show the clustering strength for a constant dark matter halo mass, making the same assumption as in our study, i.e.\ a step-function HOD with the \citet{Tinker2010} relation modeled with \texttt{halomod}. The dashed lines in the right panel indicate the duty cycle corresponding to a constant integrated quasar lifetime $t_{\rm QSO}$. 
  }
  \label{fig:dutyfig}
\end{figure*}


We obtain duty--cycle measurements ranging from $f_{\mathrm{duty}} \simeq 2\%$ to $7\%$, increasing with redshift. These values correspond to an integrated quasar lifetime of $t_{\rm QSO}\sim10^8$~years across all redshift bins. 
Note that these uncertainties represent only the statistical component obtained by forward--propagating the errors on $r_0$ and $M_{\min}$ into the halo number densities and Poisson error for $N_{\rm QSO}$. In practice, the duty cycle is subject to larger systematic uncertainties arising from assumptions about the halo--occupation distribution, the scatter between halo mass and quasar luminosity, and the choice of HOD parametrization \citep[e.g.][]{Shankar2010,Shankar2013,Pizzati2024}.

Our measurements for each redshift bin are listed in Table~\ref{tab:Mmin_fduty} and shown in Figure~\ref{fig:dutyfig}. 



\section{Summary and Discussion}
\label{sec:discussion}

Having presented our measurements of the projected correlation functions, dark matter halo masses, and duty cycles, we now turn to their interpretation. In the following, we discuss our results and compare them with previous studies. In this work, we have measured the projected correlation function $w_p(r_p)$ for $\sim\!1.3$ million quasars from the \textit{Gaia}--\textsc{Quaia} catalog across four redshift bins spanning $0 \leq z \leq 4$. Using these measurements, we inferred the characteristic host-halo masses and corresponding duty cycles, enabling a direct comparison with earlier clustering analyses.

Figure~\ref{fig:dutyfig} summarizes our results and shows the redshift evolution of the quasars' clustering lengthscale and duty cycle. 

\textbf{Clustering amplitude.} We observe that the correlation length $r_0$ increases steadily with redshift, from $r_0 \approx 6.8\,h^{-1}\,\mathrm{Mpc}$ at $0 \leq z < 1$ to $r_0 \approx 13.9\,h^{-1}\,\mathrm{Mpc}$ at $3 \leq z < 4$. At $z>3.5$ we obtain $r_0\approx 19\,h^{-1}\,\mathrm{Mpc}$, which is lower than previous work at similar redshifts \citep[][]{Shen2007, Shen2010}, as shown in Fig.~\ref{fig:shen}, and thus suggests less biased host halos. It is unclear where the discrepancy arises given that the quasars' redshift range and luminosities are similar. Differences might be due to challenges in modeling the selection function, which is more tractable with Quaia compared to SDSS. This is supported by the fact that our results and those of \citet{Shen2007} are comparable when \textit{all} quasar fields are taken into account in their analysis instead of only their \textit{good} fields, suggesting that selection effects might play a role. On the other hand the Quaia sample has less accurate redshift estimates due to the low spectral resolution of the BP/RP spectra and likely contains some contaminants, as discussed in \S~\ref{sec:data}, which could also lead to biases.  
At lower redshifts, our measurements are slightly higher than those of \cite{Eftekharzadeh2015} and \cite{White2012}, but consistent with \cite{Ross2009}.


The smooth, monotonic rise we observe with redshift appears to continue up to $z\approx 6.25$, spanning over $13$ Gyr of cosmic history. We find no evidence for any rapid change in clustering strength, despite the current lack of measurements at $4\lesssim z\lesssim 6$. At $z>7$, the measurement by \citet{Schindler2025} suggests a sudden drop in scale length and thus a potentially non-monotonic evolution. However, this study was reporting the cross-correlation of only two faint quasars with their surrounding galaxies 
and thus the results are likely highly affected by cosmic variance. 

\textbf{Host halo masses.} When translated to halo bias using the \citet{Tinker2010} relation, our measurements imply a nearly constant characteristic halo mass of $\log_{10} M_{\min}\simeq12.7$–$12.9\,h^{-1}M_\odot$ across $0 \leq z \leq 4$. These values fall squarely within the ballpark established by previous studies, and show remarkably little evolution with redshift. This near constancy confirms results from previous quasar clustering analyses \citep[e.g.,][]{Croom2005,Shen2007,Shen2010,Arita2023,Eilers2024}, and supports the idea that quasar activity preferentially occurs in halos of similar mass, independent of epoch. 
In the halo-occupation framework, this suggests the existence of a characteristic halo mass, around where quasar triggering appears most efficient. 

Note, however, that translating $r_0\!\to\!M_{\min}$ is model dependent. For instance, despite the higher clustering amplitude measured by \citet{Shen2007}, their inferred halo masses are comparable to ours, reflecting their use of a different HOD, i.e.\ the \citet{Sheth2001} bias model and \citet{Jenkins2001} mass function. These studies predict higher bias values at fixed halo mass than the \citet{Tinker2010} calibration adopted here, leading to systematically lower $M_{\min}$ estimates for a given $r_0$. 
Similarly, at $z\approx 6.25$, \citet{Eilers2024} used a step-function HOD to model both the host halos of quasars as well as those of the surrounding galaxies in the quasars' environments, while \citet{Pizzati2024} modeled the same data taking into account the quasars' luminosities and their luminosity-dependent halo mass relation. However, both approaches find host halo masses that agree within $<0.1$~dex.

\textbf{Duty cycle and lifetime.} 
We find quasar duty cycle measurements of $f_{\rm duty}\!\sim\!2$--$7\%$ increasing mildly with redshift (see right panel of Fig.~\ref{fig:dutyfig}), which 
%
suggests an integrated quasar lifetime of $t_{\rm QSO}\sim10^8$\,yr across cosmic time. This alignment is non-trivial: it means that the \emph{integrated} luminous time per halo required to match the observed quasar abundance at fixed $M_{\min}$ is roughly constant from $z\!\sim\!0$ to $z\!\sim\!4$. Physically, a lifetime of $t_{\rm QSO}\!\sim\!10^8$\,yr is comparable to a few ``Salpeter'' timescales, or $e$-folding timescales of the black hole for radiatively efficient, Eddington-limited growth, as expected in our standard picture of black hole growth \citep[e.g.,][]{YuTremaine2002}. Read together, these points favor a self-regulated black hole growth picture in which luminous accretion can proceed in one long or multiple shorter accretion episodes \citep{Eilers2017, Khrykin2019, Morey2021} whose total integrated duration is approximately $\sim10^8$\,yr per halo across cosmic time.

\begin{acknowledgments}
The authors would like to thank Yue Shen, David W. Hogg, Joe Hennawi, and Mahlet Shiferaw for helpful discussions. 

M.G.M. acknowledges the support of the MIT Summer Research Program (MSRP).

\end{acknowledgments}


\bibliography{refs}{}
\bibliographystyle{aasjournalv7}

\end{document}